\documentclass{elsart3p}

\usepackage{natbib}
\usepackage{graphicx}
\newcommand{\aap}    {A\hbox{\rm \&}A\ }
\newcommand{\aaps}   {A\hbox{\rm \&}AS\ }
\newcommand{\apj}    {ApJ\ }%
\newcommand{\apjs}   {ApJS\ }

\newcommand{\jgr}    {J.\ Geophys.\ Res.\ }

\newcommand{\nat}    {Nature\ }

\newcommand{\philtrans} {Phil.\ Trans.\ Roy.\ Soc.\ Lond.\ }
\newcommand{\solphys}{Solar Phys.\ }
\newcommand{\ssr}    {Space Sci. Rev.\ }
\newcommand{\sci}    {Science }

\newcounter{IonCS}
\newcommand{\ion}[2]{\setcounter{IonCS}{#2}#1\,{\small{\Roman{IonCS}}}}
\newcommand{\sect}[1]{Sect.\,\ref{#1}}
\newcommand{\fig}[1]{Fig.\,\ref{#1}}

\newcommand{\NNN}[1]{{{#1}}}

\def\figwidth{76mm}
\begin{document}

\begin{frontmatter}



\title{Modeling the (upper) solar atmosphere including the magnetic field%
\thanksref{label1}
}
\thanks[label1]{\normalsize\sf Accepted for publication in\newline\emph{Advances in Space Research, 21.03.2007}\newline}

\author{H.\ Peter}
\address{Kiepenheuer-Instutut f{\"u}r Sonnenphysik, Freiburg, Germany}
\ead{peter@kis.uni-freiburg.de}

\begin{abstract}

The atmosphere of the Sun is highly structured and dynamic in nature.
From the photosphere and chromosphere into the transition region and
the corona plasma-$\beta$ changes from above to below one, i.e.\
while in the lower atmosphere the energy density of the plasma
dominates, in the upper atmosphere the magnetic field plays the
governing role --- one might speak of a ``magnetic transition''.
Therefore the dynamics of the overshooting convection in the
photosphere, the granulation, is shuffling the magnetic field around
in the photosphere.
This leads not only to a (re-)structuring of the magnetic field in
the upper atmosphere, but induces also the dynamic reaction of the
coronal plasma e.g.\ due to reconnection events.
Therefore the (complex) structure and the interaction of various
magnetic patches is crucial to understand the structure, dynamics
and heating of coronal plasma as well as its acceleration into the
solar wind.

The present article will emphasize the need for three-dimensional
modeling accounting for the complexity of the solar atmosphere to
understand these processes.
Some advances on 3D modeling of the upper solar atmosphere in
magnetically closed as well as open regions will be presented
together with diagnostic tools to compare these models to
observations.
This highlights the recent success of these models which in many
respects closely match the observations.

\end{abstract}

\begin{keyword}
Sun: atmosphere \sep magnetic field

\end{keyword}

\end{frontmatter}

% main text

%=============================================================================
\section{Introduction}
%=============================================================================

The solar atmosphere extends from the photosphere and chromosphere
through the transition region into the corona.
In the photosphere and lower chromosphere, where the Fraunhofer absorption
lines are formed, the plasma is usually dominating the magnetic
field, which is frozen-in.
Models of the upper part of the convection zone and the photosphere
have to account not only for the interaction of the plasma and the
magnetic field (magneto-convection), but also for the radiative
transfer \citep[e.g.][]{Mihalas+Mihalas:1984}.
Higher up in the atmosphere the plasma becomes optically thin and the
radiative transfer \NNN{reduces to a radiative loss
function}, however, other processes such as heat conduction
\NNN{become of importance}.

What is common for the description of (almost) all interesting
phenomena on the Sun is the interaction of the magnetic field and the
plasma.
For many models one assumes the magnetic field only to provide a flow
channel, e.g.\ in penumbral filaments or coronal loops.
Despite the great success of many of these one-dimensional models
one has to account properly for the complex magnetic structure to
understand the true nature of the solar atmosphere, and this implies
to \NNN{advance} to more complex 3D models.

Traditionally the 3D magnetic structure of the outer atmosphere has
been described by extrapolations of the magnetic field, from simple
potential field models to very complex and computationally intensive
force-free extrapolations \NNN{\citep[e.g.][]{Schrijver+al:2006}.}
Such models are of vital importance to understand the large scale
structure, but are also very helpful when, e.g., investigating how
the magnetic field is rearranging itself.
During the dynamic phases of the reconnection process
\NNN{\citep{Buechner:2006}}, however, the field certainly becomes
non-force-free, and thus one also needs a model combining the
magnetic field and the plasma, i.e., a MHD model.

Due to the advancement of computer power such 3D MHD models of
structures of the solar upper atmosphere became available during
recent years.
Of course, these models cannot describe individual structures, as
e.g.\ a coronal loop, in such great detail as 1D models, but they can
properly account for the interaction and coexistence of different
structures, which is of great importance when investigating a highly
structured object such as the solar atmosphere.

Also on global scales the interaction of magnetic fields and the
plasma is of great importance.
For example, in their global models \cite{Lionello+al:2005}
investigate the large scale evolution of the (outer) corona and its
connectivity to heliosphere 
\NNN{\citep[see also the review of][]{Linker+al:2005}} .
They show that always different footpoints are connected to field
lines going into the heliosphere, which is due to on-going
reconnection.
The dynamic reconfiguration processes cannot be properly handled by a
sequence of magnetic field extrapolations.
Opening up of loops, reconnection of open regions to form loops or
interchange reconnection between open and closed regions does happen
all the time.
Such scenarios have been sketched in the past, of course, but only
now global MHD models could show that they indeed operate.

The present paper concentrates on the upper atmosphere of the Sun,
i.e.\ effects within the  \NNN{photosphere} and chromosphere as well as on
global scales will not be discussed.
The aim of this paper is to discuss coronal models for some
structures, namely the closed field structures such as (moderately)
active regions, and open structures like coronal holes.

\begin{figure*}[t]
\centerline{\includegraphics[width=0.8\textwidth,clip=true]{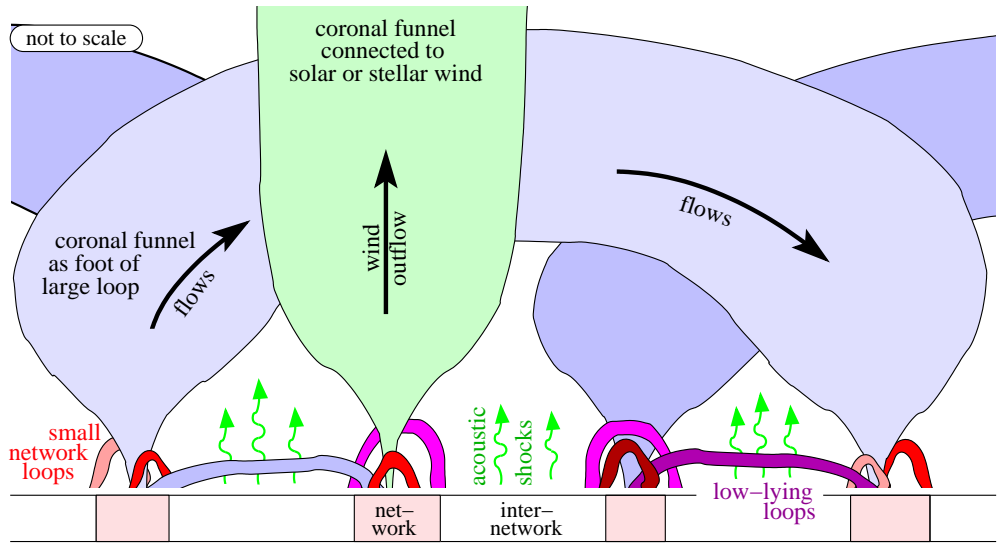}}
\caption{%
Sketch of the structure of the low corona of the Sun. This cartoon
stretches some 60\,Mm in the horizontal and 40\,Mm in the vertical
direction. It indicates the small loops within the network magnetic
patches, low-lying loops crossing network cells as well as larger
loops rooted in stronger magnetic patches and open coronal funnels.
Following \cite{Peter:2001:sec}.
\label{F:sketch}}
\end{figure*}

\NNN{%
For illustrative purposes the structure of the low corona is sketched
in \fig{F:sketch}. 
It shows the co-existence of closed magnetic structures of various
sizes: small loops with lengths below 5\,Mm connecting magnetic
patches within the chromospheric network, low-lying loops with length
below 20\,Mm spanning across network cells and finally larger loops
connecting larger magnetic patches over large distances.
The base regions of these large loops might be funnel-type, just like
the magnetically open coronal funnels.
The latter should be most prominent in coronal hole, but also might
be present in the quiet predominantly magnetically closed corona.
}

The paper will start with an overview of how to actually measure
coronal magnetic fields (\sect{S:obs.mag}), even though we know only
little about this currently.
In the following \sect{S:loops} it will be outlined how observations
(and models) require to \NNN{advance} from 1D to 3D models and why one
really needs 3D models (\sect{S:why}).
The main part of the paper, \sect{S:box}, will discuss 3D coronal box
models and the vacuum ultraviolet (VUV) spectra synthesized from
them, which allows a detailed comparison to observations.
Open structures, especially the origin and acceleration of the fast
wind in coronal holes will be the subject of \sect{S:open}, before 
\sect{S:conclusions} concludes the paper.

%%%-overview-%%%%%%%%%%%%%%%%%%%%%%%%%%%%%%%%%%%%%%%%%%%%%%%%%%%%%%%%%%%%%%%%
%%%%%%%%%%%%%%%%%%%%%%%%%%%%%%%%%%%%%%%%%%%%%%%%%%%%%%%%%%%%%%%%%%%%%%%%%%%%%

%=============================================================================
\section{Measurements of coronal magnetic fields}            \label{S:obs.mag}
%=============================================================================

Before discussing  modeling of the solar atmosphere including the
magnetic field, the attention should be drawn on how to actually
observe the magnetic field in the corona.
There are various techniques providing the coronal magnetic field.

Direct coronagraphic \NNN{measurements} of the Stokes vector e.g.\ in the
infrared lines of \ion{Fe}{13} at 10747\,{\AA} and 10798\,{\AA}
applying the Zeeman effect show magnetic fields in bright loops being
of the order of 20 Gauss \citep{Lin+al:2000}.
Such measurements can be applied only for regions seen above the
limb, of course, and are most valuable for the large-scale structure
of the corona.
Smaller areas, especially low in the corona, are not accessible through
this technique, because they are occulted by the design of the
coronagraph.

Radio observations allow to investigate the coronal magnetic field
also on the disk, e.g.\ through the Zeeman effect \citep{White:2005}.
Radio interferometers such as VLA provide a remarkable spatial
resolution of 12\,arcsec, and experiments for the future FASR facility
showed that the magnetic field strength could be retrieved in a
reliable way \citep{White:2005}.
The problems here are the detailed interpretation that the emission
at a given frequency does originate from a complex volume in the
corona, and that these observations are of value mainly for
active regions.

Another valuable tool are infrared observations in the \ion{He}{1}
(10830\,\AA) triplet using Zeeman and Hanle effect giving magnetic
fields for emerging flux regions \citep{Lagg+al:2004}.
However, \ion{He}{1} is not a coronal line and can thus only indicate
magnetic fields in just newly emerging flux regions, which still carry
cool plasma.

Extrapolation techniques based on photospheric and/or chromospheric
measurements are also a very valuable tool to investigate coronal
magnetic fields.
\NNN{As a rule of thumb one might trust these extrapolation on
scales typically comparable to or larger than a fraction of the
chromospheric network, i.e., some 5--10\,Mm.
On smaller scales still the expansion of the magnetic field into the
chromosphere, i.e.\ the equilibrium of magnetic and gas pressure, can
be expected to dominate.}
However, whenever interesting events occur, such
extrapolations, no matter how fancy, can not be trusted completely,
as then the assumption of a force free state is most likely
violated.
\NNN{While in such events the overall field structure on the larger
  scales mentioned above might not change dramatically, the
  force-free violation might be more severe on smaller scales in a
  transient manner.
  This will be certainly true on scales on which explosive events do
  occur \citep[i.e.\ some Mm;][]{Innes+al:1997} as well as on scales
  not yet resolvable, e.g. in the case of nanoflares.}
Thus such extrapolations are important and helpful to understand the
overall structure and long-term evolution, but real \NNN{measurements} of
the \NNN{coronal} magnetic field \NNN{with a spatial resolution of
  1\,Mm and better} are desperately needed in order to pin-point the
relevant processes during phases when the heating mechanism is
showing itself through a dynamic non-force-free event, \NNN{as in
  flares, explosive events or nanoflares.}
\NNN{While the above mainly concerns closed field regions, the same
  is also true for magnetically open region as found in coronal holes.
Following the idea of e.g. \cite{Axford+McKenzie:1997} magnetic
reconnection of open field with bipolar network flux is the energy
source for the heating of the open corona and the wind acceleration.
It would be of high interest to detect the non-force-free magnetic
reconfiguration predicted by this furnace model through the direct
measurement of the magnetic field.}

The ideal tool to investigate the magnetic field in the corona is to
use lines which are formed in the transition region and corona, i.e.\
to use the same VUV emission which is currently widely used to
investigate the corona with imagers such as TRACE or EIT or
spectrometers such as SUMER or CDS.
Of course, because of the complicated spatial structure of the corona
and the high variability, the interpretation of the polarized light
in the VUV will not be possible in a simple and straightforward way,
and is thus confronting us with the same problems as radio observations.
The coronal models, which become more and more elaborate and
realistic, will be of pivotal importance for a proper interpretation
of VUV and radio spectro-polarimetric data.

The importance of UV and VUV spectro-polarimetry from space has been
emphasized by \cite{Trujillo-Bueno+al:2005}, who suggest to use the
Lyman series of hydrogen for an investigation of the Hanle effect.
Another approach is to measure the Stokes vector in strong lines such
as the \ion{C}{4} doublet at 1548\,{\AA} and 1550\,{\AA}, which is
formed in the transition region.
This is suggested by \cite{West+al:2005}, who are currently building
an instrument for a series of rocket flights, the first probably in
spring 2007 for a test of the system (Solar Ultraviolet Magnetograph
Investigation, SUMI).
Because of the limited count rate during the rocket flight, they
expect to get a signal only in Stokes V, which would be in itself a
great step forward, if they succeed.

In the future we have to aim at a combination of  spectro-polarimetry
in the radio and the VUV.
In the radio such techniques already exist, but they have to be
refined, also with respect to spatial resolution.
For VUV spectro-polarimetry we will certainly have to wait some
time until \NNN{space instrumentation} will be available, but this is
the instrument to aim at when investigating the magnetic structure of
the corona.

%=============================================================================
\section{From 1D to 3D: the closed corona}                     \label{S:loops}
%=============================================================================

The \NNN{most conspicuous} basic building block of the corona is the
coronal loop.
Observations show that more or less semi-circular shaped loops are
the most prominent structure in the corona, and they are found in
active regions, after flares, in quiescent regions in the network,
etc.
Therefore most 1D models for the solar corona dealt with loop models,
one of the most well-known being the RTV models
\citep{Rosner+al:1978}.

Even though it is not clear what the ``micro-structure'' of an
observed loop would be, i.e.\ if it consists of many individual
strands or if it is a bundle of parallel field lines
\NNN{\citep{Klimchuk:2006}}, many loop models exist solving the mass,
momentum and energy balance along a loop-shaped 1D structure, some
now with adaptive mesh refinement and even a self-consistent
treatment of the ionization and radiation
\citep[e.g.][]{Bradshaw+Mason:2003b,Mueller+al:2003}.
Such loop models allow a very detailed description of the thermal and
dynamic properties within the loop (or strand) modeling complex
non-linear processes, e.g.\ catastrophic cooling
\citep{Mueller+al:2004}, which is also found in observations
\citep{Schrijver:2001}.
\NNN{1D models for open structures will be discussed in \sect{S:oneDwind}.}

The main disadvantage of these 1D loop models is that they (usually) are
not able to treat the heating process in a physical way.%
\footnote{%
\NNN{This is not true for the 1D corona and wind models of e.g.\
  \cite{Marsch+Tu:1997} or \cite{Tu+Marsch:1997} as discussed in
  \sect{S:oneDwind}.}
}
Since the early RTV work almost all 1D models assume a parameterized
form of the distribution of coronal heating, e.g.\ exponentially
decaying with height, i.e.\ not really incorporating a physical
heating mechanism into the model.
However some models made an attempt to include a distribution of heat
into the 1D models as found from other studies.
%% \citep[e.g.][]{Walsh+al:????}.

In the mid 1970ies the Skylab observations showed that the
chromospheric network is expanding to higher temperatures
\citep{Reeves:1976}, which lead to assume a funnel-type structure of
the transition region and low corona \citep{Gabriel:1976}.
However, it became clear later that these funnels \NNN{\emph{alone}}
are not a good representation of the corona, \NNN{even though they are
an important ingredient (cf. \fig{F:sketch})}.
For example in the \NNN{magnetically closed} corona the emission measure
$EM=\int_V\,n_e^2\,{\rm{d}}V$, i.e.\ a measure of the emissivity of
the corona at a given temperature, could not be reproduced by the
funnel models.
Because of the ineffective heat conduction at low temperatures the 1D
models give an extremely thin transition region, i.e.\ a very steep
temperature gradient, and thus in the model there is not enough
material at low temperatures to account for the observed increase of
the emission measure towards lower temperatures below some $10^5$\,K.
\cite{Dowdy+al:1986} proposed that loops at various temperatures
could account for that increase and through this introduced a
complicated magnetic structure of the low corona, sometimes also
called ``magnetic junkyard'' \NNN{(cf. \fig{F:sketch})}.
Later it was proposed that signatures in the spectra of VUV emission
lines support this scenario \citep{Peter:2000:sec}.
\NNN{%
Furthermore there is ample observational evidence from VUV line
emission maps acquired with SUMER that low-lying loops with lengths
below 10--20\,Mm do exist \citep[e.g.][]{Feldman+al:2003}.
}
However, all these scenarios are not really based on a physical
model, but are merely trying to find a plausible way to draw a sketch
of the magnetic structure of the low corona.

\NNN{%
In open field regions such as coronal holes it seems problematic that
the above scenario for the (predominantly) closed field regions
holds.
There other processes might be more important.
Recently \cite{Esser+al:2005} have shown that the (fast) solar wind
outflow of the plasma along funnel type structures might well result
in significant violation of the ionization equilibrium by lifting a
sizable fraction of neutral material higher up into hotter regions of
the funnel.
This can then produce an increased emission in Ly-$\alpha$ which
agrees well with observation.
However, it still has to be explored to which extent this process will
predict the observed emission from the transition region, i.e. of
lines formed at a couple of 100\,000\,K.
}

\NNN{While large-scale structures such as a (polar) coronal hole or
an active region complex seem to be rather stable for days and weeks,
on smaller scales of below 10\,Mm the corona is highly dynamic.
When analyzing the spatial magnetic structure on these smaller scales
this variability is adding to the complexity of the problem.  
For example,} when modeling an explosive event, it is usually assumed
that the ``background'' magnetic structure remains the same, while
only a small perturbation is shuffling around the magnetic field
leading to a reconnection event \NNN{\cite[e.g.][]{Innes+Toth:1999}.}
When investigating observations with imagers or
spectrographs, one sees a very strong variability in time indicating
that the ``natural'' state of the the transition region and corona \NNN{on
scales below 10\,Mm} is a dynamic one \NNN{\citep{Innes:2004}.}

%=============================================================================
\section{Why 3D models?}                                         \label{S:why}
%=============================================================================

One could argue that a composite model consisting of a large number
of loops \NNN{and open structures} would be sufficient to describe
the solar corona.
\NNN{Furthermore one could argue to neglect the magnetically
open structures in these models as they provide only a minor
contribution to the emission in X-rays or the far VUV, i.e.\ in the
wavelength bands accessible to Yohkoh, or the EIT or TRACE coronal
channels.}
As  the corona and transition region are built up by (loop-like)
magnetic structures which are rooted in the solar photosphere, one
could use an extrapolation of the magnetic field to identify the
loops and then investigate the resulting structure.

For certain problems this is certainly a good approach.
For example \cite{Schrijver+al:2004} did a global extrapolation of
the observed photospheric magnetic field to define the loops and used
a simple scaling law for the flux of mechanical energy $F_H$ into the
loop, $F_H\propto\,B^\beta\,L^\lambda\,f$.
Here $B$ is the magnetic field at the base of the structure, $L$ its
half length, $f=\exp[-(B/500\,G)^2]$ a factor accounting for
reduced heating in sunspots, and $\beta$ and $\lambda$ are free
parameters.
Each loop was then described in a static 1D model.
The authors then calculated how this constructed multi-loop corona would
appear in an X-ray observation (Yohkoh SXT) and compared it to the
real observation.
By varying $\beta$ and $\lambda$ they found that a heating scaling
with $\beta\approx{1.0}{\pm}0.5$ and $\lambda\approx{-0.7}{\pm}0.3$
would give the best fit.
Even the best fit is not really representing the real corona, which
is not surprising as it is based on a very simple model.
Nevertheless it can be used to estimate how
the energy input into the corona is roughly scaling with the magnetic
structures.
Most important, this kind of model opens the possibility for more
realistic global models of stellar corona, which are not accessible
to direct observations.

For other purposes one-dimensional models, or such multi-loop models
would not be sufficient.
For example \cite{Schrijver+Title:2003} showed that depending on the
distribution of internetwork magnetic flux only some 50\% of the
coronal magnetic field above the quiet Sun does come from network
magnetic patches.
This implies that the simple idea of a coronal loop being rooted in a
single (or even a few) patches of strong magnetic flux in the
photosphere is not applicable.
Furthermore \cite{Jendersie+Peter:2006} could show that based on
current instrumentation we are not even able to determine where the
coronal field is really rooted, \NNN{on a scale of a fraction of a
  super-granule (say 5\,Mm)}.
For only small differences in the distribution of weak internetwork flux
elements one gets radically different connections from the
photosphere into the corona.
Therefore when describing the corona in more detail on a scale
corresponding to \NNN{a fraction of} a single active region or the
chromospheric network,
one has to account for the complex magnetic connectivity, because a
simple picture of a coronal structure rooted in a single (or few)
strong magnetic patch(es) does not really apply.
On top of that the problem is even more complicated, because the
distribution of magnetic flux in the photosphere changes constantly
on time scales from minutes \NNN{(granules, ${\approx}$1\,Mm)} to days
\NNN{(super-granules, ${\approx}$20\,Mm)}, which is shorter than the 
lifetime of many loops.
\NNN{Loop arcades, e.g.\ as seen by TRACE, overall last longer (many
days) than the nano-flaring processes associated with them, so they
appear to be rigid, or solid.
However, the magnetic connectivity within the loop
structures might constantly change while the loop is sitting
seemingly still, i.e.\ the many strands of the loop could constantly
change their identity.
Currently it is not clear if, and if yes to what extent, this affects
the appearance of the loop as a whole.}
Therefore we might ask ourselves the question what sense the concept
of a coronal loop as a rigid magnetic loop really makes.

In consequence, to describe the structure of the transition region
and corona, one has to account for the changing ``boundary
conditions'' in the photosphere and include a proper treatment of the
3D structure of the magnetic field.

%=============================================================================
\section{3D MHD coronal box models}                              \label{S:box}
%=============================================================================

One of the most important ingredients for a coronal model is a proper
energy equation.
\NNN{Because of the strong flows observed in the active regions as
  well as quiet corona, the enthalpy flux can play a major
  role.
And only if the energy balance, including furthermore the}
energy input, heat conduction and radiation
are treated well enough, the density of the corona will be correct (as
good as possible...).
This is because basically the energy input sets the pressure of the
coronal structure: for larger energy input more energy will have to
be radiated in the transition region.
When the heating rate increases this consequently leads to
(chromospheric) evaporation, while the coronal temperature changes only
little as the heat conduction acts as a thermostat through the strong
dependency on temperature (${\propto}T^{5/2}$).
Thus only a proper energy equation will assure proper densities and
consequently proper velocities within the computational domain.

\NNN{%
A severe problem is that the energy dissipation processes are
occuring on scales much smaller than currently observable or
resolvable with the MHD-type models to describe e.g.\ an active region
complex or a coronal loop.
Thus in terms of kinetic physics it is questionable is a ``proper
energy equation'' in fluid terms can even be defined
\citep[e.g.][]{Marsch:2006}.
However, one might (or have to) hope that when describing the corona
with a resolution of the current MHD models (some km in 1D, some 100
km in 3D), the energy deposition as a function of space and time
is comparable to the result a micro-physical model smoothed to the
MHD scales would give.
Also the resistivity in the MHD models is basically set by the
(spatial) resolution they can achieve.
Due to the limitations for the magnetic Reynolds number this is a
common property for all MHD-type models.
However MHD studies with different spatial resolutions and
resistivities showed that the energy deposition does not change
significantly over a range of scales (for resolution and resistivity)
accessible to MHD experiments
\citep{Galsgaard+Nordlund:1996,Hendrix+al:1996}.
Future work will have to show to what extent the MHD approach is a
good representation for the real Sun, but currently it is the most
appropriate (and only) tool to describe structures such as active
region systems or loop complexes.
}

The basic test for every coronal model has to be whether it is
matching with observations.
The most straight-forward way for such a test is to compare
observable quantities such a line intensities or shifts synthesized
from the model to real observations, as this eliminates the many
problems of inversions of VUV spectral data
\citep[e.g.][]{Judge+McIntosh:1999}.
In the following a 3D MHD coronal box forward model will be discussed
which compares very well with observations concerning e.g.\ the
emission measure, line shifts or temporal variability.

%-----------------------------------------------------------------------------
\subsection{3D forward box model}                             \label{S:threeD}
%-----------------------------------------------------------------------------

A 3D MHD model presented by
\cite{Gudiksen+Nordlund:2005a,Gudiksen+Nordlund:2005b} includes the
atmosphere from the photosphere to the lower corona in a
60${\times}$60\,Mm horizontal times 37\,Mm vertical box.
It solves the mass, momentum and energy balance, the latter
including classical heat conduction \citep{Spitzer:1956} and
optically thin radiative losses as piecewise power laws.
The temperature of the chromosphere is kept near a prescribed profile by
Newtonian cooling.
Initially the magnetic field is given by a potential field extrapolation
with the lower boundary as observed by MDI/SOHO for an active region.
Further on in the simulation the field evolves self-consistently and
becomes non-potential, of course.
The system is driven by horizontal motions at the lower boundary.
The flow field is constructed using a Voronoi-tessellation technique
\citep{Okabe+al:1992} and reproduces the typical pattern of the granular
motions of the Sun \citep{Schrijver+al:1997}.
By this also the power spectra of the velocity and the vorticity are
reproduced.
\NNN{%
The spacing of the (non-uniform) grid goes down to 150\,km
and is as good it can be given the size of the structure modeled here
(60\,Mm).
Of course all the shortcomings of MHD models as discussed above apply
here.
For the upper boundary condition it is assumed that the magnetic
field above the computational domain is potential, which, of course,
prevents the development of larger open magnetic structures (and
was not the aim of that study). 
This has to be kept in mind when discussing the Doppler shifts of
coronal lines (\sect{S:diagnostics}) and the apparent lack of hot
blueshifted material, i.e.\ a wind.
}

This procedure leads to a heating of the corona just in the way
\cite{Parker:1972} suggested.
The field line braiding gives high coronal temperatures of a million K, and
a system of hot loops is forming, connecting the magnetic concentrations
of the active region.
Furthermore the system reaches some sort of quasi-stationary state, with
large fluctuations in time and space, or in other words, the system is
quite dynamic.
The detailed discussion of the MHD results can be found in
\cite{Gudiksen+Nordlund:2005a,Gudiksen+Nordlund:2005b}.

Based on the above MHD model results for the \NNN{density and temperature} 
\cite{Peter+al:2004,Peter+al:2006} calculated the emissivity at each
grid point under the assumption of ionization equilibrium using the
atomic data base CHIANTI.
Assigning a spectral line profile at each grid point with a line
width corresponding to the thermal width, one can integrate the
spectra from the 3D box along a line of sight, e.g.\ the vertical.
This results in 2D maps of spectral profiles, which can be analyzed in the
same way as observations, e.g.\ through maps of line intensities or shifts.
For a detailed discussion of the synthesis of the emission lines from the
transition region and corona and the assumptions and limitations see
\cite{Peter+al:2004,Peter+al:2006}.

\begin{figure*}[t]
\centerline{\includegraphics[width=1.00\textwidth]{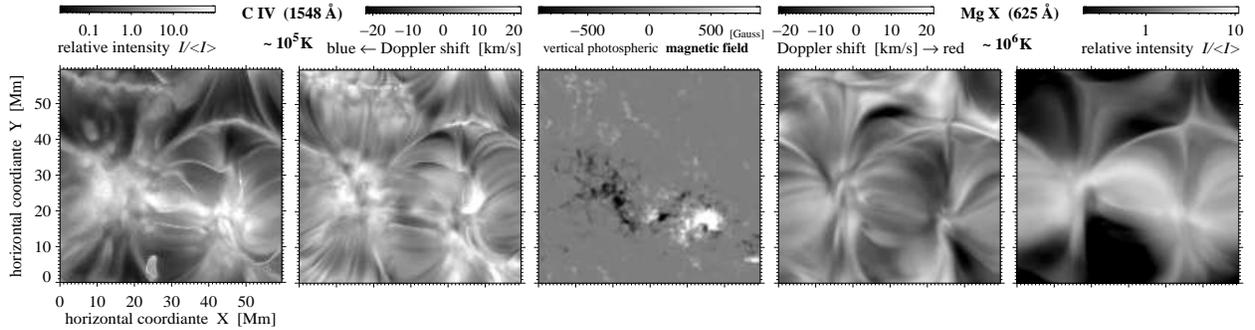}}
\caption{%
Maps in line intensity and Doppler shift in the transition region (left
two panels) and the corona (right two panels) as derived from a snapshot
of a 3D MHD coronal model.
In these maps the computational box is viewed from above, i.e.\ this
represents the situation when observing near disk center.
The middle panel shows the vertical magnetic field at the bottom of the
computational domain, i.e.\ in the photosphere.
Following \cite{Peter+al:2006}.
\label{F:images}}
\end{figure*}

Figure \ref{F:images} shows in the middle panel the vertical magnetic
field at the bottom of the computational box, i.e.\ in the photosphere.
To the left and right are maps in line intensity and line shift as derived
from the spectra computed from the MHD model.
They represent a view from the top onto the box, corresponding to an
observation at disk center.
As found with observations the spatial structures in the transition region
line \ion{C}{4} (1548\,\AA) at 10$^5$\,K are much smaller and finer than in
the corona seen in \ion{Mg}{10} (625\,\AA) formed at 10$^6$\,K.

%%%-overview-%%%%%%%%%%%%%%%%%%%%%%%%%%%%%%%%%%%%%%%%%%%%%%%%%%%%%%%%%%%%%%%%
%%%%%%%%%%%%%%%%%%%%%%%%%%%%%%%%%%%%%%%%%%%%%%%%%%%%%%%%%%%%%%%%%%%%%%%%%%%%%

%-----------------------------------------------------------------------------
\subsection{Diagnostics of the 3D model: VUV spectra}   \label{S:diagnostics}
%-----------------------------------------------------------------------------

In the following a short overview will be given how VUV spectra
synthesized from a 3D coronal box model can be used to test the
model.
It should be stressed that the model results presented in the
following show the best overall match to the observations published so
far for the differential emission measure, Doppler shifts or rms
fluctuations.

%%%-DEM-%%%%%%%%%%%%%%%%%%%%%%%%%%%%%%%%%%%%%%%%%%%%%%%%%%%%%%%%%%%%%%%%%%%%%
\begin{figure}[t]
\centerline{\includegraphics[width=\figwidth]{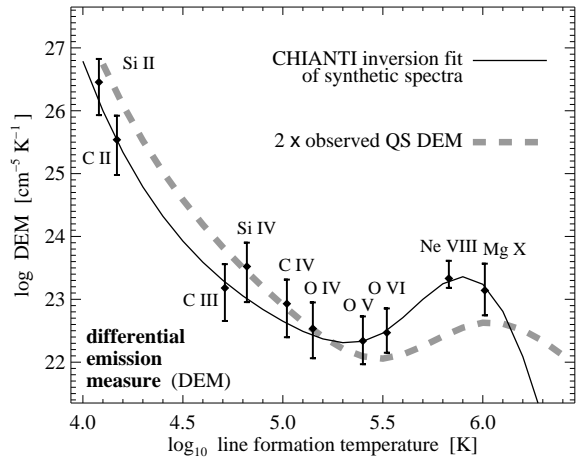}}
\caption{%
Differential emission measure (DEM) from
a 3D MHD box model compared to observations.
The solid line shows the fit from the DEM inversion based on the lines
displayed as bars, which have been synthesized from the MHD model
\NNN{(integrated over the whole box).}
The thick dashed line is based on a DEM inversion using observed quiet Sun
disk center line radiances observed with SUMER
\protect\citep{Wilhelm+al:1998} scaled by a factor of two.
Following \cite{Peter+al:2004}.
\label{F:dem}}
\end{figure}
%%%%%%%%%%%%%%%%%%%%%%%%%%%%%%%%%%%%%%%%%%%%%%%%%%%%%%%%%%%%%%%%%%%%%%%%%%%%%

\subsubsection*{Diffential emission measure}
%-----------------------------------------------------------------------------

For each time step \cite{Peter+al:2004} used the radiances of a
number of VUV emission lines synthesized from a MHD model to perform
a differential emission measure (DEM) analysis, \NNN{i.e.\ they used
  the maps of the spectral profiles (``synthetic observations'') as
  described in the previous section and integrated the line profiles over
  the whole map. Thus the line radiances represent the total emission
  of the given VUV lines integrated over the whole computational domain.}
\NNN{As with real observations one can then use an inversion procedure
  to derive the differential emission measure (DEM), e.g. with the
  help of the procedures provided with the} CHIANTI atomic data
  package \citep{Dere+al:1997,Young+al:2003}.%
  \NNN{\footnote{%
  \NNN{As with any DEM inversion, many implicit assumptions apply, e.g.\
  ionization equilibrium, constant abundances, constant pressure
  atmosphere.
  }}}
\fig{F:dem} shows the resulting DEM curve as a solid line for a single
time step derived from the intensities of a given set of lines,
\NNN{which characterizes the (average) small active region
  simulated in the MHD model}.
For comparison the thick dashed line shows the inversion of
actually observed intensities of the same lines.
The match of the model with the observations is remarkable, especially,
the 3D MHD model used in the \cite{Peter+al:2004} study
\citep{Gudiksen+Nordlund:2002} is reproducing the observed trend of
the emission measure below ${\sim}10^5$\,K, while previous (1D and
2D) models failed to predict this increase of DEM towards lower
temperatures (cf.\ \sect{S:loops}).

In the above model the increase towards the low transition region is
caused by numerous low-lying (intermittent) cool structures.
This confirms the scenario outlined by \cite{Dowdy+al:1986}, in which a
hierarchy of large and small, hot and cool loops exist in the transition
region and corona as already discussed in \sect{S:loops}.
Despite the variability of the low corona, the emission measure (averaged
over the computational domain, viz.\ the active region) changes only
little with time.

%%%-Doppler-shifts-%%%%%%%%%%%%%%%%%%%%%%%%%%%%%%%%%%%%%%%%%%%%%%%%%%%%%%%%%
\begin{figure}[t]
\centerline{\includegraphics[width=\figwidth]{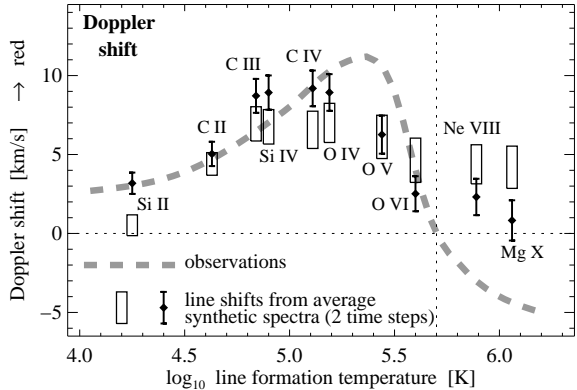}}
\caption{%
Average Doppler shifts of the synthesized spectra when viewing the
computational box from straight above for two different time steps (bars
and rectangles).
The height of the bars indicate the standard deviation of the Doppler
shifts for the respective time step.
The thick dashed line shows the trend found in observations
\citep{Peter+Judge:1999}.
Following \cite{Peter+al:2004}.
\label{F:doppler}}
\end{figure}
%%%%%%%%%%%%%%%%%%%%%%%%%%%%%%%%%%%%%%%%%%%%%%%%%%%%%%%%%%%%%%%%%%%%%%%%%%%%%

\subsubsection*{Average Doppler shifts}
%-----------------------------------------------------------------------------

Since the discovery of the systematic transition region line redshifts
by \cite{Doschek+al:1976}, the explanation of these persistent Doppler
shifts has been a challenge for modelers.
The discovery that the lines formed in the (low) corona show a net
blueshift by \cite{Peter:1999full} and \cite{Peter+Judge:1999} using
SUMER/SOHO data added a new quality to this challenge.
The trend of the net Doppler shift at quiet Sun disk center is shown in
\fig{F:doppler} as a thick dashed line \citep[data compiled
from][]{Brekke+al:1997,Chae+al:1998,Peter+Judge:1999,Teriaca+al:1999:ar}.

\NNN{In order to derive the average Doppler shift corresponding to
  disk center observations one can investigate the Doppler maps of
  the synthetic spectra with a vertical line-of-sight, i.e.\ when
  looking at the computational box from straight above (cf.\
  \fig{F:images}).
  }
\NNN{The average line shifts as computed by} \cite{Peter+al:2004} 
for two different time steps 7\,min apart are
plotted in \fig{F:doppler} as bars and rectangles (the heights of the
bars and rectangles indicating the scatter of the Doppler shifts).%
\NNN{\footnote{
\NNN{Actually, the average values for the shifts of the Doppler maps
  are quite similar to the Doppler shifts of the average spectra.}}}

The line shifts are caused by the dynamic response of the atmosphere
to the energy input due to the braiding of magnetic field lines.
The flows are partly induced through energy deposition in and
subsequent expansion of the corona, partly from evaporation of
chromospheric material due to heating at low heights and the heat
flux from the corona into the chromosphere.

Throughout the transition region \NNN{(${<}5{\cdot}10^5$\,K)} the
match of the observed Doppler shifts with those of the spectra
synthesized from the model is \NNN{quite} good.
It should be stressed here that no fine-tuning was applied, but these
Doppler shifts follow naturally from the driving of the corona through
the footpoint motions of the magnetic field lines.
In the low corona \NNN{(${>}5{\cdot}10^5$\,K)} the synthesized
spectra do not show the blueshifts as they are observed
\citep{Peter:1999full}, but this problem might be overcome (at least
partly) by a more appropriate hand\-ling of the upper boundary
condition.

%%%-rms-fluctuations-%%%%%%%%%%%%%%%%%%%%%%%%%%%%%%%%%%%%%%%%%%%%%%%%%%%%%%%
\begin{figure}[t]
\centerline{\includegraphics[width=\figwidth]{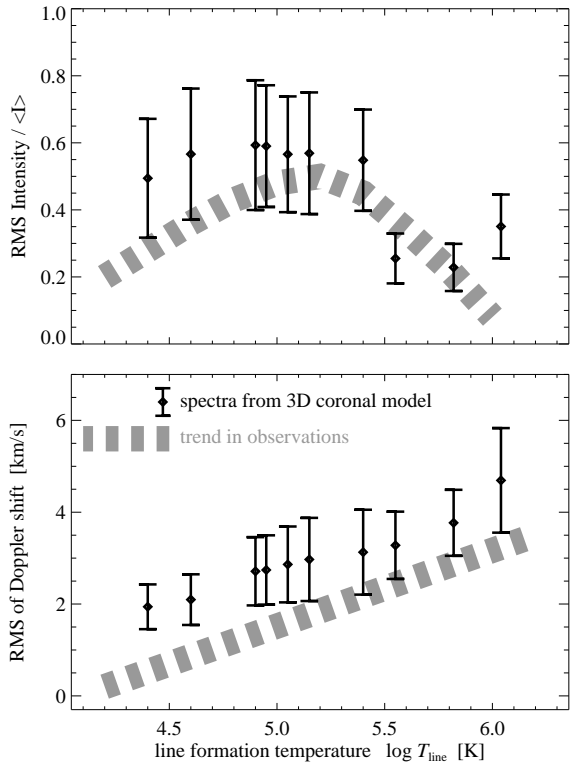}}
\caption{%
RMS fluctuations in line intensity and shift from the spectra
derived from the MHD coronal model for a number of VUV lines as a function
of line formation temperature.
The thick dashed line shows the trend found in observations by
\cite{Brkovic+al:2003}.
Following \cite{Peter+al:2005:ccmag}.
\label{F:rms}}
\end{figure}
%%%%%%%%%%%%%%%%%%%%%%%%%%%%%%%%%%%%%%%%%%%%%%%%%%%%%%%%%%%%%%%%%%%%%%%%%%%%%

\subsubsection*{Temporal variability}
%-----------------------------------------------------------------------------

Based on the synthesized maps of line intensities and shifts
generated from a line of sight integration along the vertical
direction, the rms fluctuations at each spatial pixel within the
respective map have been evaluated by \cite{Peter+al:2005:ccmag}.
Figure \ref{F:rms} shows their results for a number of VUV lines as a
function of line formation temperature.
The diamonds show the (spatially) averaged values for the rms fluctuation
for each line and the bars represent the scatter of the rms values (standard
deviation).

This is exactly the same procedure as used by \cite{Brkovic+al:2003} to
reduce their observational data from CDS and SUMER on SOHO.
The trends they found with their observations is over-plotted as thick
dashed lines in \fig{F:rms}.
It has to be emphasized that the spatial resolution of the maps in
line intensity and shift derived from the MHD model has been reduced
to match the spatial resolution of the SOHO instruments, as this has
an effect on the absolute values of the rms fluctuations (the trend
with temperature, however, does not depend on the spatial resolution).

Together with the good match between the average Doppler shifts and
emission measure from observations and the spectra computed from the
3D MHD coronal box model presented above, this is yet another strong
indication that the underlying 3D coronal model is a good description
for the coronal heating mechanism.

%-----------------------------------------------------------------------------
\subsection{The magnetic structure}
%-----------------------------------------------------------------------------

%%%-plasma-beta-%%%%%%%%%%%%%%%%%%%%%%%%%%%%%%%%%%%%%%%%%%%%%%%%%%%%%%%%%%%%
\begin{figure}[t]
\centerline{\includegraphics[width=\figwidth]{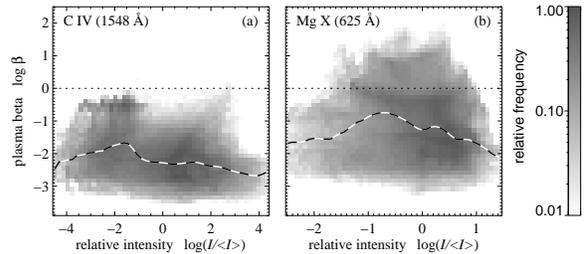}}
\caption{%
2D histograms of plasma--$\beta$ as a function of line intensity
(a: \ion{C}{4}, b: \ion{Mg}{10}).
The dashed lines show the median value of plasma--$\beta$ in a given
interval of intensity.
Please note that the 2D histograms are plotted on a log-scale.
Following \cite{Peter+al:2006}.
\label{F:beta}}
\end{figure}
%%%%%%%%%%%%%%%%%%%%%%%%%%%%%%%%%%%%%%%%%%%%%%%%%%%%%%%%%%%%%%%%%%%%%%%%%%%%%

To investigate the interaction of the magnetic field and the plasma one
can discuss plasma--$\beta$, i.e.\ the ratio of the gas pressure to
the magnetic pressure, $\beta=2\mu\,p/B^2$.
\fig{F:beta} shows the relation of $\beta$ to the synthesized
transition region (a; \NNN{\ion{C}{4}}) and coronal (b; \NNN{\ion{Mg}{10}}) emission within the
computational box as done by \cite{Peter+al:2006}.
\NNN{%
The plots show the 2D histograms of the plasma beta as a function of
(normalized) intensity of the respective line, i.e.\ the color coding
represents the fraction of the volume where one finds the respective
combination of plasma-$\beta$ and normalized intensity.
}
The median relations \NNN{between  plasma-$\beta$ and the normalized
  intensities} are over-plotted as dashed lines.
The transition region and corona are mostly in a low-$\beta$ state,
however with noticeable exceptions in the coronal part!

It is clear that in the whole transition region the assumption of a
low-$\beta$ plasma is very good (\fig{F:beta}a).
This is different, however, in the corona when relating $\beta$
to the \ion{Mg}{10} emission (\fig{F:beta}b).
A large fraction of the not-low-beta plasma in the corona
is at low densities, \NNN{i.e.\ at $\log\,I/{\langle}I{\rangle}<0$,} where the
magnetic field is very weak, too.
\NNN{As these are regions of low emission, one might be inclined to
  disregard this.}
However, even in the volume with a relative intensity larger than the
median intensity (i.e.\ $\log\,I/{\langle}I{\rangle}>0$ in
\fig{F:beta}b) contributing ${\sim}$90\% to the total coronal
\ion{Mg}{10} emission, some ${\sim}$5\% of this bright material has
values of $\beta>1$ \citep{Peter+al:2006}.
\NNN{These are regions of rather high densities but low magnetic
  field, and they can be found e.g.\ in regions along magnetic
  neutral lines across which the direction of the magnetic field
  changes and thus strong currents heat the plasma and finally cause
  the density to increase.} 

In conclusion, in significant patches of the corona, even above
active regions, $\beta$ is in fact not much less than unity.
%
%% Therefore forces such as gravity and inertia, are able to distort the
%% magnetic field to some degree.

This again emphasizes that for ``good'' coronal model it is not
enough to only extrapolate the magnetic field and then solve a lot
of 1D loop-like problems along each field line.
Instead, one has to account for the interaction of the plasma and the
magnetic field in a more complex 3D model.

%=============================================================================
\section{3D MHD models for open structures}                     \label{S:open}
%=============================================================================

So far the discussion concentrated on magnetically closed structures,
such as active regions or the chromospheric network.
In magnetically open regions the plasma is not trapped by the
magnetic field, but can escape along magnetic field lines.
This is the basic reason why coronal holes, i.e.\ the source region
of the fast solar wind, appear dark.
Because a large part of the energy goes into acceleration of the
wind, there is less energy to heat the plasma.
Consequently the temperature is a bit lower than in the quiet Sun,
typically below $10^6$\,K \citep{Wilhelm+al:1998corona}.
The effect of the reduced heating rate is much stronger on the
pressure, which results in a lower density in the corona above the
holes --- we see less emission from the coronal holes.

%-----------------------------------------------------------------------------
\subsection{1D scenarios and models for the fast wind} \label{S:oneDwind}
%-----------------------------------------------------------------------------

As for the quiet and active Sun structures, which are dominated by
loops, the first models for the open corona, which is the same as for
the fast wind, were 1D models stretching from the base of the corona
into interplanetary space, with an area expansion factor accounting
for the (super) radial expansion of the magnetic field.

As with the loops these models used a parameterized form of the energy
input, which was typically concentrated close to the Sun.
As mentioned earlier, in order to properly understand the mass loss,
especially to describe the mass loss rate self-consistently set by
the energy input, one has to include a transition region to account
for the heat flux back to the Sun
\citep{Hammer:1982a,Hammer:1982b,Withbroe:1988}.
Only then one can describe properly the interplay between coronal
heating, densities, temperatures and solar wind acceleration, as done
first by \cite{Hansteen+Leer:1995}.
However, as for the 1D loop models \NNN{(cf.\ \sect{S:loops})}, these
models did not include a physical heating mechanism.

\NNN{In contrast,} \cite{Axford+McKenzie:1997} put forward an idea
that small-scale reconnection in the magnetic concentrations of the
chromospheric network (i.e.\ between open and closed field lines)
leads to high-frequency Alfv\'en waves which propagate upwards into
the corona, \NNN{a concept also called magnetic furnace.}
These waves can then resonantly interact with the protons and heavy
ions in the open corona and drive the wind as has been described by
\cite{Marsch+Tu:1997} and \cite{Tu+Marsch:1997}.
Furthermore one can also account for the expansion of the magnetic
field directly above the magnetic concentrations in the chromosphere
forming large funnels \NNN{\citep{Marsch+Tu:1997:chromo,Hackenberg+al:2000}.}

These 1D models including a physical mechanism for the heating and
acceleration of the plasma describe a situation where the material is
continuously and constantly accelerated from the chromosphere out into
the solar wind \NNN{\citep[e.g.][]{Holzer:2005}}.
By this they implicitly assume that the magnetic configuration from
the upper chromosphere into the solar wind is rather stable!
This would strongly contradict the scenario outlined for the closed
field regions in the previous section, where the magnetic field in
the photosphere is constantly shuffled, resulting in a continuous
change also higher up.

%-----------------------------------------------------------------------------
\subsection{Observing the source of the fast wind}
%-----------------------------------------------------------------------------

It has been known for a while that the fast wind is originating from
coronal holes, and observations of blueshifts in coronal lines within
coronal holes confirmed this \citep{Rottman+al:1982,Orrall+al:1983}.
Even the transition region lines of \ion{C}{4} showed a higher
fraction of blueshifted regions in coronal holes than in the quiet
Sun \citep{Dere+al:1989,Peter:1999full}.

The strongest blueshifts are found in the darkest coronal regions
\citep{Wilhelm+al:2000} and they seem to be concentrated
at intersections of network lanes \citep{Hassler+al:1999}.
A comparison of the Doppler shift patterns to the photospheric
magnetic field showed that these strong blueshifts, interpreted as
the base of the outflow, coincide with strong magnetic concentrations
in the network \citep{Xia+al:2004}.
These data seemed to support the wind scenario described above,
namely that funnel-type structures emerging from strong network
magnetic patches define the flow channels of the wind reaching out
into the interplanetary medium.

\NNN{One can investigate the magnetic connectivity from the
  photosphere to the corona in coronal holes and compare this to the
  Doppler shift signals of transition region and (low) coronal lines.
  This shows that} the blueshifts in coronal lines coincide with the
  magnetic funnels, but that at those locations no systematic
  blueshifts can be found in transition region lines \NNN{within the
  instrumental uncertainties} \citep{Tu+al:2005}.
Based on this observation \cite{Tu+al:2005} argued that the plasma of the solar
wind outflow has to be injected somewhere between the formation
heights of the transition region and coronal lines they inspected
(\ion{C}{4} and \ion{Ne}{8}), i.e.\ somewhere around 10\,Mm.
If this is correct, it changes dramatically our view of the formation
of the wind at its very base, \NNN{because it would challenge the old
  paradigm that there is a continuous outflow of the wind from the
  chromosphere into the heliosphere}.
\NNN{However, because of the higher densities in the transition region
the expected outflow speed of the wind should be rather low
($\approx$1\,km/s) at the level where one can expect \ion{C}{4} to
form.
This is just at the resolution limit of current instrumentation,}
e.g.\ SUMER/SOHO can determine Doppler shifts down to some
1--2\,km/s \citep{Peter+Judge:1999}.
\NNN{Therefore it remains unclear, whether there is a continuous
  outflow from the chromosphere, with a small not yet detectable
  velocity at \ion{C}{4} heights, or if the velocity in \ion{C}{4} is
  really zero and the wind is injected between the levels of
  \ion{C}{4} and \ion{Ne}{8} as} suggested by \cite{Tu+al:2005}.

%%%-open-field-model-%%%%%%%%%%%%%%%%%%%%%%%%%%%%%%%%%%%%%%%%%%%%%%%%%%%%%%%
\begin{figure}[t]
\centerline{\includegraphics[width=\figwidth]{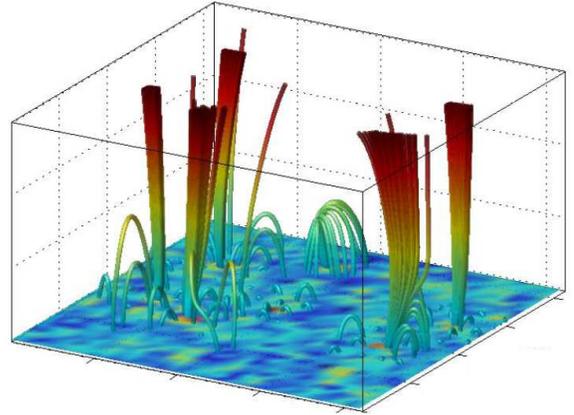}}
\caption{%
Initial setup of the magnetic field in the 3D box model for an open
field region by \cite{Buechner+Nikutowski:2005}.
The bottom boundary of the magnetic field is taken from observations
and is driven by whirl flows, which brings together the open and
closed regions leading to reconnection.
From \cite{Buechner+Nikutowski:2005}.
\label{F:open}}
\end{figure}
%%%%%%%%%%%%%%%%%%%%%%%%%%%%%%%%%%%%%%%%%%%%%%%%%%%%%%%%%%%%%%%%%%%%%%%%%%%%%

%-----------------------------------------------------------------------------
\subsection{Accounting for the magnetic structure: 3D models for the
  onset of the fast wind}
%-----------------------------------------------------------------------------

To investigate the scenario proposed by \cite{Tu+al:2005} a
numerical experiment was conducted by
\cite{Buechner+Nikutowski:2005}.
They start with a magnetic configuration given at the lower
boundary from observations and then drive the system by whirl flows
imposed on the lower boundary.
Through this they bring together open and closed magnetic field
structures which results in reconnection.
Their model contains a plasma--neutral gas coupling, i.e.\ it
includes effects beyond MHD. 
For the reconnection they used a switch-on-resistivity, i.e.\ the
resistivity is non-zero only where the current-carrier-velocity
(${\propto}j/n$) is above a certain threshold.
By this their treatment of the reconnection is more detailed than the
3D box models for the closed corona discussed in \sect{S:threeD},
however, \NNN{they used a simplified energy equation, not accounting
  for heat conduction or energy losses through optically thin
  radiation \citep[cf.][]{Buechner+al:2005}.}
The magnetic configuration (at the initial state) is shown in
\fig{F:open}.

As a consequence of driving the magnetic field through the whirls
at the bottom, the reconnection between the open and closed regions
injects energy and plasma into the open funnels and through this
drives a wind out of the funnels.
Roughly, the plasma below $10^5$\,K falls down, plasma above
$10^5$\,K leaves the computational box through the top.
The downdrafts are mostly concentrated above the magnetic
concentrations at the bottom boundary.
Thus this simulation supports the general picture suggested by
\cite{Tu+al:2005}.
However, further modeling with a proper inclusion of an energy
equation and also an investigation of the VUV emission to be expected
from this model is needed before a final conclusion can be drawn.

At the moment we have to conclude that even in the very quiet coronal
holes, the assumption of a stationary magnetic field configuration
simply channeling the solar wind outflow is not justified.
To really understand the origin of the solar wind we have to account
for the complex interaction between different magnetic field structures
and also between the magnetic field and the plasma.

%=============================================================================
\section{Conclusions}                                    \label{S:conclusions}
%=============================================================================

A realistic model of the (upper) solar atmosphere has to account for
the complexity of its magnetic structure.
Together with the constant rearrangement of the magnetic field at
photospheric levels this leads to an ongoing energy release in the low
parts of the transition region and corona and through this drives the
highly dynamic upper atmosphere.

The current box models for magnetically closed regions as discussed in
\sect{S:box} show a \NNN{qualitatively} good match to observed
quantities, such as emission measures, Doppler shifts of
rms-variability.
This presents ample evidence that the heating through field line
braiding is a prime candidate to heat the corona.
\NNN{Of course, other means of energy input into the corona
  might also reproduce the average quantities of e.g.\ emission
  measure, Doppler shift or line widths as presented in this paper.
  It remains to be seen if other models based on different heating
  processes will give the same or different results, once they are
  pushed to produce observables as now done for the field line
  braiding model.}
In the future more efforts have to be undertaken not
only to increase the spatial resolution of the simulations, but also to
actually include a physical mechanism for the dissipation of magnetic
energy during the reconnection events.

The box models for magnetically open region, i.e.\ for the onset of
the fast wind (\sect{S:open}), have not yet been analyzed in such detail
concerning observable properties, but they support some recent
interpretation of observational data suggesting that plasma is
injected in funnel-type structures to form the fast solar wind.

Together with the very successful global coronal models, which have
only briefly been touched upon in this paper, now complex
three-dimensional models become available accounting properly for the
plasma and the magnetic field as well as their interaction.
We might hope that the coming years will bring further advances of
these models and their diagnostic capabilities. \NNN{Thus we will improve
our understanding of} the structure, dynamics and heating of the upper
atmospheres of the Sun and solar-like stars.

%\bibliography{bib/all,bib/new,bib/local,bib/cospar}
%\bibliographystyle{elsart-harv}

\end{document}